MEASUREMENTS ON THE KINETIC PHYSICS IN STREAMER DYNAMICS

F.S. Mozer [1,2], K.-E. Choi[1], R. Sydora[3], A. Voshchepynets[4]
[1]Space Sciences Laboratory, University of California, Berkeley, 94720, USA.
[2]Physics Department, University of California, Berkeley, 94720, USA.
[3]Physics Department, University of Alberta, Edmonton T6G 2E1, Alberta, Canada.
[4]Dept. of System Analysis and Optimization Theory, Uzhhorod National University, Uzhhorod, Ukraine.

**ABSTRACT**

A major goal of solar physics is understanding the transition of the medium from the closed-loop magnetic configuration of the corona to the open structure of the heliospheric current sheet. The evolution of solar wind streamers, an essential component of this transition, has been observed in-situ for the first time by measuring a pass through a streamer stalk at 11.7 solar radii. A plasma rest frame DC electric field, reaching an amplitude of 400 mV/m, was observed as the magnetic field changed from +1100 nT to -1100 nT. This electric field violates the frozen-in condition so it must be understood via the non-MHD, Hall **J×B**/ne term of the Generalized Ohm's Law. Two plasma regimes containing the large electric field were observed. In the first regime, the magnetic field was large, the ion beta was small (~0.1), the ratio of the ion skin depth to ion gyroradius was large (>10), and the current of ~0.3 mA/m$^2$ was carried by electrons. In the second regime, the magnetic field was small, the ion beta was large (~10), the ion skin depth was equal to the ion gyroradius, and the ~10 mA/m$^2$ current was carried by the ions. Also, measurements during 15 such crossings showed that the plasma flow speed inside the current sheets exceeded that outside six times and the 326 km/sec average speed inside the current sheets exceeded the 266 km/sec outside. Such findings challenge the traditional consensus that streamers are the source of the "slow" solar wind.

## INTRODUCTION

The complex magnetic topology of the solar corona is dominated by structures such as helmet streamers which are believed to be primary conduits for the release of the slow and intermediate solar wind. Understanding the plasma dynamics within these structures is critical for resolving their evolution into the heliospheric current sheet (HCS)—the fundamental transition from the closed-loop magnetic environment of the corona to the open, radial structure of the heliosphere. This transition follows a well-described but physically enigmatic topological sequence driven by the expansion of the solar wind.

At the base of the corona, a helmet streamer consists of closed magnetic loops that trap dense plasma. Further outward—at a height of roughly three solar radii—the magnetic pressure of these closed loops is no longer sufficient to contain the thermal pressure of the coronal plasma and the field lines pinch together at a null point known as the cusp. Beyond the cusp, the magnetic field lines are stretched into a nearly radial orientation by the accelerating solar wind. This thin, high-density region of oppositely directed magnetic fields forms the streamer stalk, the precursor to the HCS. As the Sun rotates, this sheet evolves into a warped, three-dimensional surface that separates the two large-scale magnetic polarities of the heliosphere.

In the heliosphere, these large-scale fields and currents are often interpreted through the lens of magnetohydrodynamics (MHD) theory [Dey et al., 2026; Lionello et al, 2026]. In this approach, the opposite-polarity radial magnetic field lines on the two sides of the solar equator are supported by a required longitudinal current. This current interacts with the radial magnetic field to produce a meridional **JxB** force that is balanced by the meridional pressure gradient associated with the meridional variations in plasma density and temperature.

While fluid-based descriptions [Raissi et al, 2018], including advanced closures like the Physics-Informed Neural Network (PINN) approach [Cuomo et al, 2022] successfully model global dynamics, they remain approximations that struggle to capture the non-ideal microphysics of the interior regions. Because such non-MHD descriptions are not well-understood, it is imperative to directly measure the electric fields and currents to resolve the kinetic processes occurring in these regions. In this paper we describe the electric fields, magnetic fields, and currents measured in 15 streamer structures observed by the Parker Solar Probe during orbits 16 through 23, at radial distances of 11.7 to 17 solar radii in order to begin exploring these kinetic processes.

## DATA

The electric and magnetic field data on the Parker Solar Probe were obtained by the FIELDS instruments [Bale et al, 2016]. Their data are presented in the spacecraft frame, for which the X-axis points eastward in the equatorial plane, the Y-axis points southward, and the Z-axis points Sunward. Because all 15 crossed streamer structures yielded similar results, the crossing on orbit 21 is studied in detail in this paper. Figure 1a gives the radial component of the magnetic field measured during the two-hour streamer crossing while Figures 1b and 1d present the measured X and Y electric field components (the Z-component was not measured) while Figures 1c and 1e present the X- and Y- components of **-vxB,** the cross product of the negative solar wind velocity and the magnetic field. For MHD to be valid, **E** must be equal to **-vxB**. As seen in panels 1b and 1c this is not the case for this streamer crossing because the measured $E_x$ was huge, (as large as 400 mV/m, approximately the largest DC electric field observed during the entire Parker Solar Probe mission) and $-(vxB)_x$ was approximately equal to zero. However, $E_y$ was approximately equal to $-(vxB)_y$, the differences being attributable to uncertainties due to the short antenna length of three meters [Mozer et al, 2019] and shadowing of the plasma detectors by the spacecraft body. Thus, the X-components of Figure 1 show that the spacecraft was not in an MHD regime and the Y-components provide validation of instrumental performance.

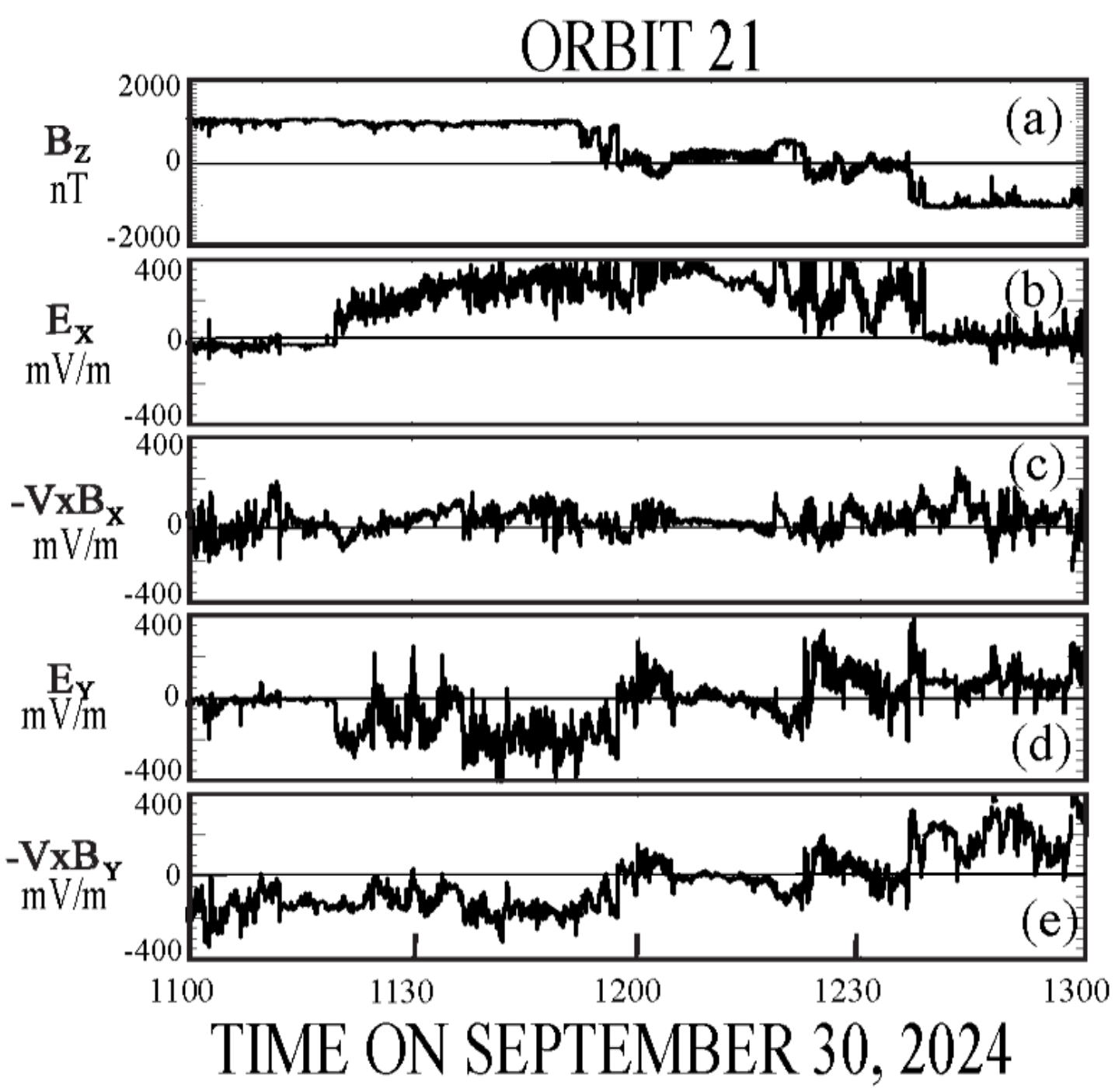

Figure 1. Comparison of Measured and Convective Electric Fields during a Streamer Crossing. (1a) The radial magnetic field showing its rotation across the current sheet. (1b, 1d) Components of the electric field in the spacecraft frame. (1c, 1e) Negative components of the convective electric field. The massive amplitude discrepancy between panels (1b) and (1c) identifies the non-MHD nature of the encounter, while the agreement between (1d) and (1e) validates the instrument performance.

Figure 2 presents plasma parameters observed during traversal of the extended streamer stalk, showing rotation of the radial magnetic field in 2a. Inside this extended streamer stalk the solar wind flow of 2b increased from about 250 to 450 km/sec, which is opposite from expectations of the streamer being the source of the slow solar wind. Panel 2c shows that the plasma density increased in the stalk, as expected. Panel 2d presents the X-component of the electric field in the plasma rest frame (called the prime frame), in which $E'_x = E_x + (vxB)_x$. The electric field in the prime frame does not participate in an **E**x**B** drift because that component of the electric field was removed when transforming into the prime frame. Instead, the prime electric field is related to other plasma parameters by the Generalized Ohm's Law (GOL), which is

$$\mathbf{E'} = \mathbf{E} + \mathbf{v} \times \mathbf{B} = \eta \mathbf{J} + \mathbf{J} \times \mathbf{B}/ne - \nabla \mathbf{P}/ne + (me/\ ne^2)\ \partial \mathbf{J}/\partial t \qquad (1)$$

where
**v** = plasma bulk velocity
η = collisional resistivity
**J** = current density
n = plasma density
**∇P** = pressure gradient

For further analyses, the unmeasured $E_Z'$ was reconstructed by assuming that the parallel electric field was zero at large scales, or that

$$E_Z' = (-E_X'B_X - E_Y'B_Y)/B_Z \tag{2}$$

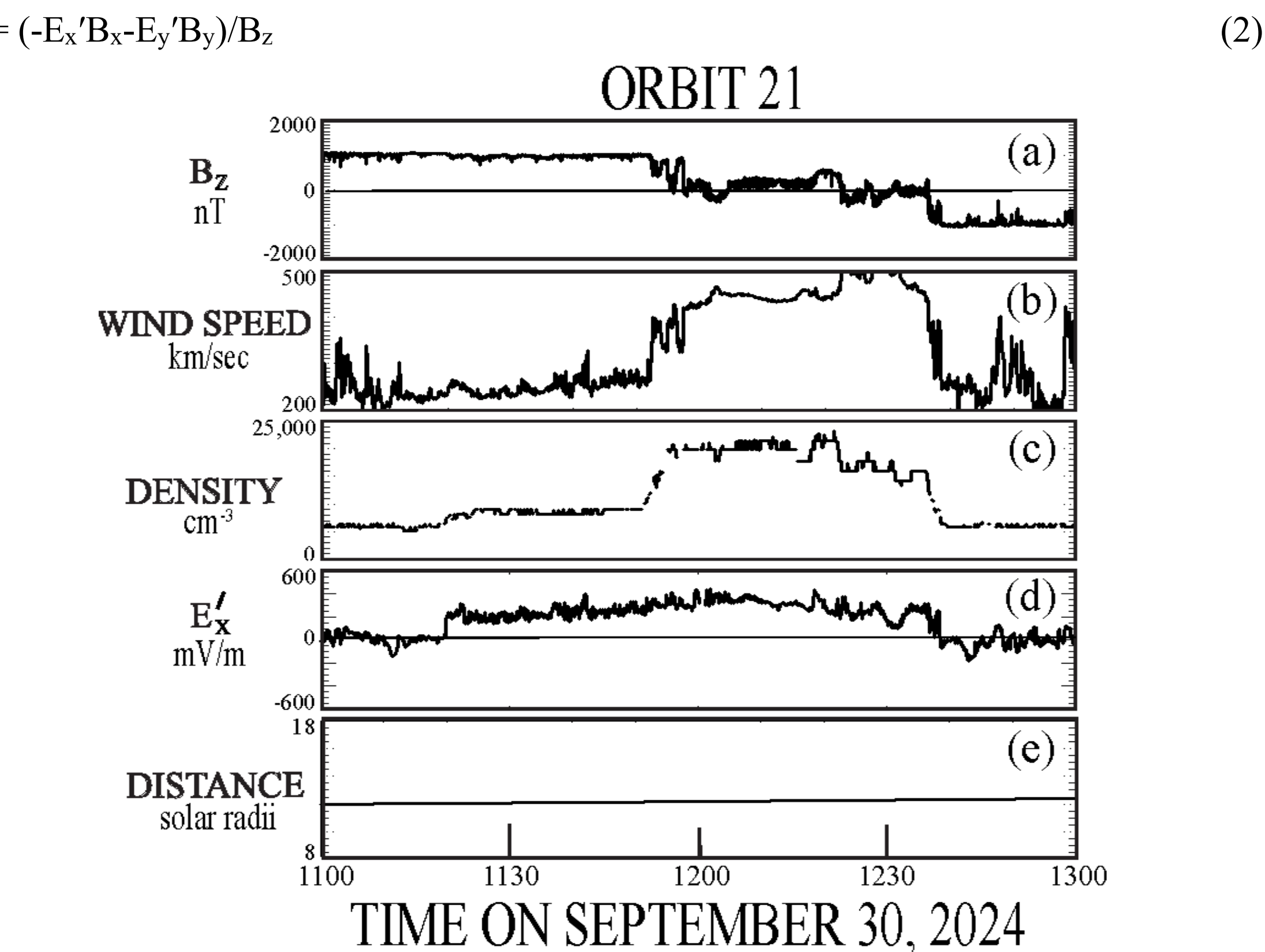

Figure 2. Plasma Parameters and Velocity Profiles in the Extended Streamer Stalk. (2a) Radial magnetic field illustrating its rotation during the two-hour traversal. (2b) Solar wind bulk flow velocity, showing an anomalous increase from 250 to 450 km/s within the stalk. (2c) Plasma density profile demonstrating the characteristic density enhancement of the streamer structure. (2d) The X-prime component of the electric field as transformed into the plasma rest frame, which provides the primary input for the Generalized Ohm's Law (GOL) analysis. (2e) The radial distance of the spacecraft from the Sun.

With these additions and corrections, the three components of the prime electric field are presented in Figure 3. This must be a local rather than global description of the electric field since a non-zero, longitudinal or X-component electric field around the Sun would require an unimaginable time rate of change of the magnetic field. As noted earlier, understanding and explaining this electric field requires non-MHD theory.

To interpret the electric field in terms of the GOL, the relative amplitude of each of its terms must be known. In general, the Hall term in the solar wind is thought to be the largest term by more than an order-of-magnitude [Priest and Forbes, 2000]. However, the magnitudes of the terms depend on the plasma parameters and the measurements of interest were made in an unusual environment. If the Hall term is the largest term, **J** may be determined because all other terms in the equation $\mathbf{E}' = \mathbf{J}\times\mathbf{B}/ne$ are known. The result, illustrated in Figures 4b, 4c, and 4d, shows that, in the local non-MHD region, the main current flowed in the meridional or Y-direction. This is a different flow direction from that further out in the MHD region where the flow is mainly in the longitudinal or X-direction to create the heliospheric current sheet. That the current density is extremely turbulent at times in Figure 4 is a result of both the uncertainties in the denominator of the current calculation, which is near zero at times, and because the current is truly turbulent.

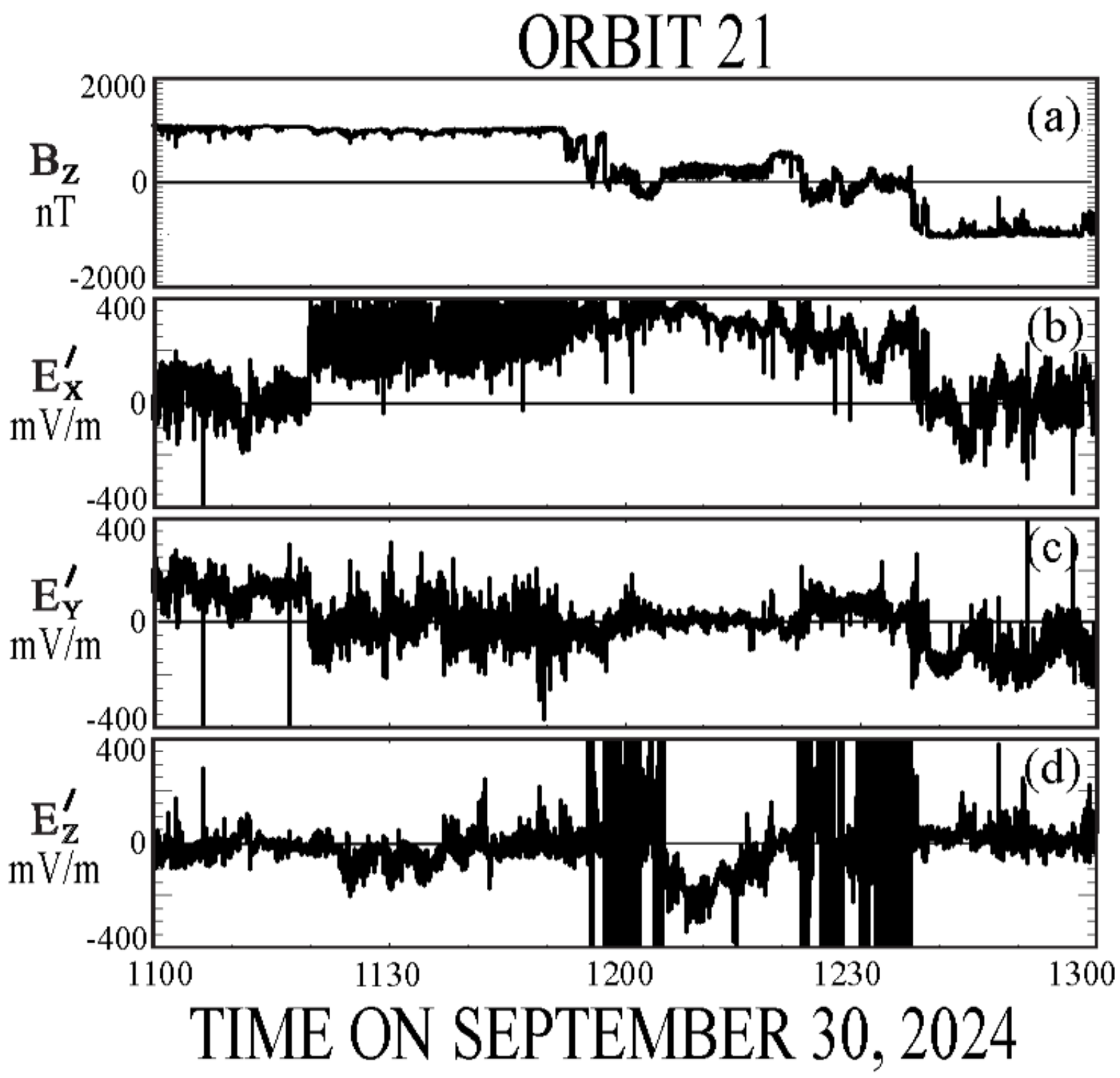

Figure 3. Three-Component Reconstruction of the Rest-Frame (Prime) Electric Field in panels 3b, 3c, and 3d. The persistence of non-zero components serve as a local signature of kinetic-scale processes that cannot be explained by global MHD equilibrium.

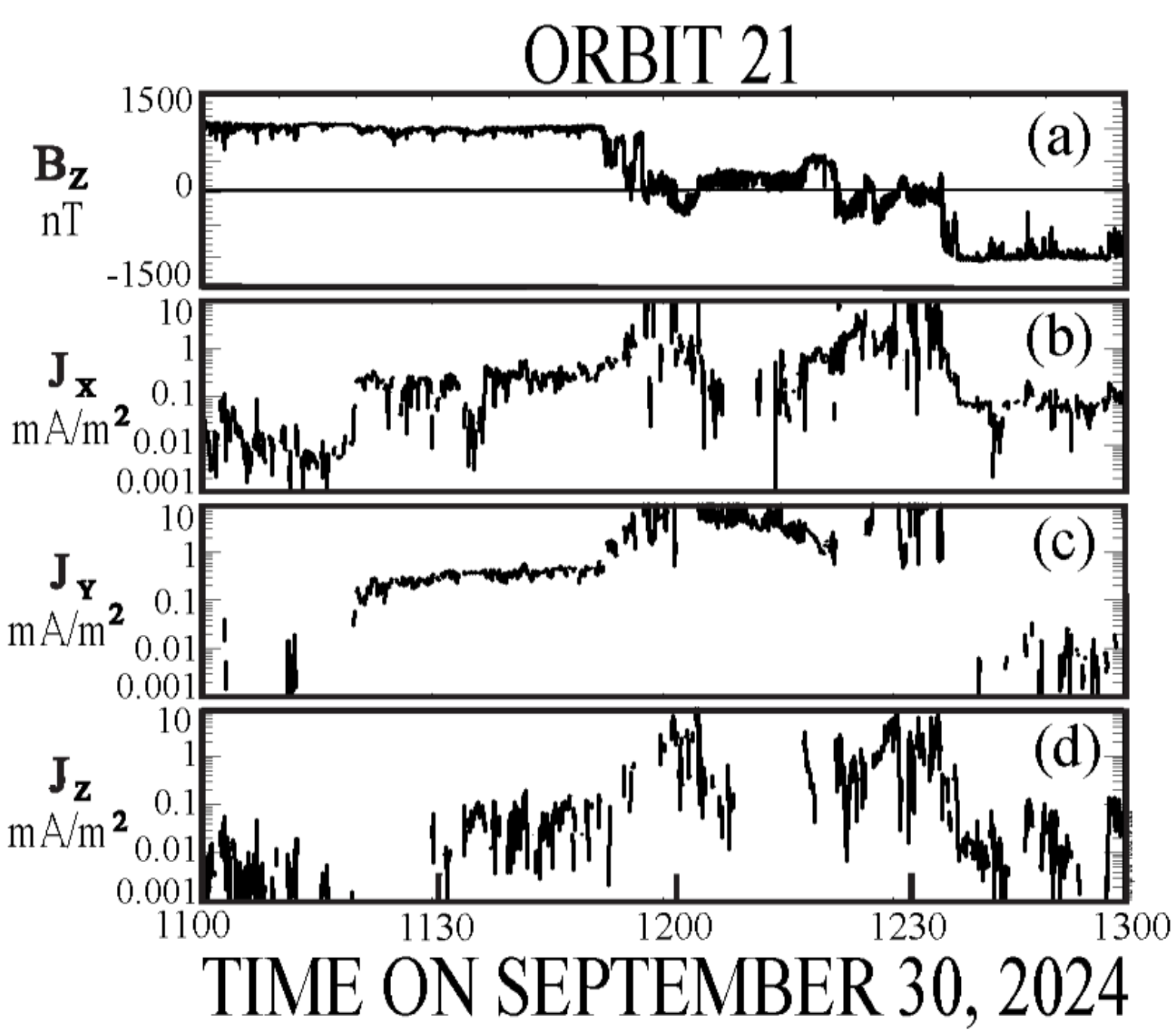

Figure 4. Panels 4b, 4c, and 4d give the three components of the current density derived from the Hall term of the GOL during the crossing of the current sheet indicated by the magnetic field of 4a. The dominant meridional flow represents the kinetic precursor to the global heliospheric current sheet

**DISCUSSION**

The large electric field of Figure 5a may be divided into the two intervals of 11:20-11:50 and 11:50-12:40, which are bounded by the dashed vertical lines. In the first interval, the magnetic field is large so the ion gyroradius is small and the skin-depth-to-gyroradius is large (~15) while the ion beta is small (<<1). In the second interval, the magnetic field is small, so the skin-depth-to-gyroradius is ~1 and beta ~10.

The interval from 11:20 to 11:50 represents a classic Sub-Ion/Hall Regime. The combination of a very low plasma beta and a high skin-depth-to-gyroradius ratio indicates that the ions were inertially decoupled from the magnetic field. At this scale, the ions were "too heavy" to track the rapid magnetic fluctuations and steep gradients of the structure. Consequently, the electrons—which remained firmly "frozen" to the field lines due to their much smaller mass and skin depth—carry the majority of the current. The resulting Hall current (roughly 0.5 mA/m$^2$) is driven by the "slippage" between these stationary ions and the magnetized, flowing electrons

In contrast, the interval from 11:50 to12:40 transitioned into a High-Beta/Thermal Regime as the spacecraft crossed the null point in the magnetic field. With a plasma beta significantly exceeding unity and a skin-depth-to-gyroradius approaching unity, the physical constraints on the ions change fundamentally. In this "soft" magnetic environment, the plasma thermal pressure dominated the magnetic energy density, and the ions were no longer inertially decoupled at the scale of the current structure. Here, the ions participated fully in the plasma dynamics, and the current density was driven by ion pressure gradients and diamagnetic drifts. This explains the order-of-magnitude increase in current density, which peaks near 10 mA/m², marking the transition from an electron-only kinetic carrier to an ion-dominated bulk current.

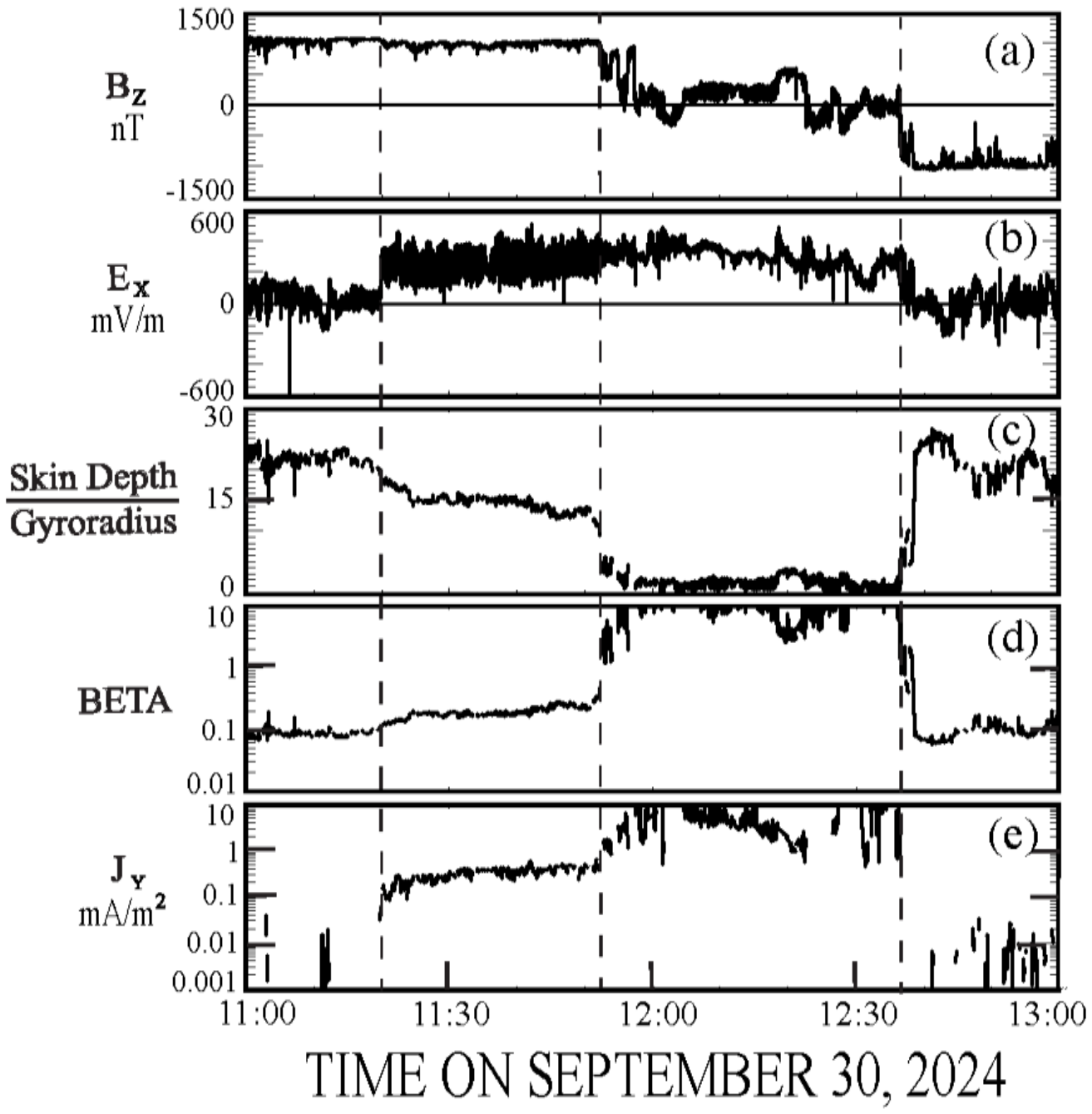

Figure 5. Magnetic field (5a) during the crossing of the current sheet. (5b) the non-zero plasma frame electric field that identifies this as a non-MHD region. The skin depth divided by the ion gyroradius (5c) and plasma ion beta (5d). (5e) One component of the current density. The large electric field region is divided into two regimes bounded by the dashed vertical lines.

In Figure 2b, it was shown that the solar wind velocity of 450 km/sec inside the current sheet greatly exceeded the 250 km/sec velocity outside. This example differs from earlier conclusions that streamers are the source of the slow solar wind [Abbo et al, 2016]. In the present data set, the solar wind flow speed inside the current sheet exceeded that outside the sheet in six of the 15 examples and was smaller in four examples. Also, the 326 km/sec average speed inside the current sheet exceeded the average speed of 266 km/sec outside the region. Such in-situ results question the concept that the slow solar wind is generated in streamers.

In global models, a 15 solar radius location is often used as the "hand-off" point where coronal physics (heating and acceleration) transitions into pure heliospheric propagation. That this happens in the discussed example is shown by the presence of non-MHD prime electric fields in Figure 3 and non-MHD currents in Figure 4 that are controlled by the ratio of the ion skin depth to the ion gyroradius.

## SUMMARY

This paper provides a detailed empirical look at the kinetic-to-fluid transition of the solar wind within 17 solar radii. By examining 15 streamer crossings, we have moved beyond the approximations of global fluid models to resolve the actual electric fields and current densities at the streamer base. The data show:

- Non-Ideal Physics: We measured electric fields in the plasma frame that are incompatible with standard MHD theory because they are as large as 400 mV/m. By considering the ratio of the ion skin depth to the ion gyroradius and the ion beta, we distinguished between regions in which the electrons carried the current and regions in which ions carried the current.
- Velocity Anomalies: Our data show that the solar wind speed inside the streamer stalk was frequently faster than the wind outside it. With an average internal speed of 326 km/s versus an outside speed of 266 km/s, this velocity inversion questions the concept that the slow solar wind is generated in streamers.

## ACKNOWLEDGEMENTS

This work was supported by NASA contract NNN06AA01C. The authors acknowledge the extraordinary contributions of the Parker Solar Probe spacecraft engineering team at the Applied Physics Laboratory at Johns Hopkins University. The FIELDS experiment on the Parker Solar Probe was designed and developed under NASA contract NNN06AA01C.